\documentclass[journal=jacsat,manuscript=article]{achemso}

\usepackage[version=3]{mhchem} 
\usepackage{rotating}       
\usepackage{subcaption}
\usepackage{pdflscape}

\makeatletter
    \ifXeTeX%
        \newcounter{rotfigcount}%
        \patchcmd{\@xrotfloat}%
            {\begin{minipage}\textheight}%
            {%
                \begin{minipage}\textheight%
                \stepcounter{rotfigcount}%
                \label{rotfig:\therotfigcount}%
                \ifnumodd{\getpagerefnumber{rotfig:\therotfigcount}}%
                    {}%
                    {}%
            }%
            {}{}%
    \fi%
    \ifLuaTeX%
        \newcounter{cntsideways}%
        \AtBeginShipout{%
            \ifnum\zref@extractdefault{rotate\number\value{page}}{page}{0}=0%
                \PLS@RemoveRotate%
            \else%
                \ifnumodd{\thepage}{\PLS@AddRotate{90}}{\PLS@AddRotate{-90}}%
            \fi}%
        \patchcmd{\@xrotfloat}%
            {\begin{minipage}\textheight}%
            {\begin{minipage}\textheight\rotatesidewayslabel}%
            {}{}%
        \newcommand\rotatesidewayslabel{\stepcounter{cntsideways}%
        \zlabel{tmp\thecntsideways}\zlabel{rotate\zref@extractdefault{tmp\thecntsideways}{page}{0}}}%
    \fi%
\makeatother

\author{Kay S. Schaller}
\email{kaysc@dtu.dk}
\affiliation[DTU Bioengineering]
{Department of Biotechnology and Biomedicine, Technical University of Denmark}
\alsoaffiliation[DTU Kemi]
{Department of Chemistry, Technical University of Denmark}
\author{Jeppe Kari}
\affiliation[DTU Bioengineering]
{Department of Biotechnology and Biomedicine, Technical University of Denmark}
\author{Kim Borch}
\affiliation[Novozymes]
{Novozymes A/S}
\author{G\"unther H.J. Peters}
\affiliation[DTU Kemi]
{Department of Chemistry, Technical University of Denmark}
\author{Peter Westh}
\affiliation[DTU Bioengineering]
{Department of Biotechnology and Biomedicine, Technical University of Denmark}

\title[Binding prediction of multi-domain cellulases with a dual-CNN]
  {Binding prediction of multi-domain cellulases with a dual-CNN}

\abbreviations{
CAZy: Carbohydrate-Active Enzyme,
CBH: Cellobiohydrolase,
CBM: Cellulose-Binding Module,
CD: Catalytic Domain,
Cel: Cellulase,
CNN: Convolutional Neural Network,
EG: Endoglycanase,
Grad-CAM: Gradient-weighted Class Activation Mapping
GH: Glycoside Hydrolase,
LIE: Linear Interaction Energy (method),
MAE: Mean-Absolute Error,
MD: Molecular Dynamics,
ML: Machine Learning,
MSE: Mean-Squared Error,
RMSE: Root-Mean Squared Error,
SGD: Stochastic Gradient Descent,
\textit{Tr}: Trichoderma reesei,
XAI: Explainable Artificial Intelligence,
}

\begin{document}

%
%
%
%
%

\begin{abstract}
Cellulases hold great promise for the production of biofuels and biochemicals. However, they are modular enzymes acting on a complex heterogeneous substrate. Because of this complexity, the computational prediction of their catalytic properties remains scarce, which restricts both enzyme discovery and enzyme design. Here, we present a dual-input convolutional neural network to predict the binding of multi-domain enzymes. This regression model outperformed previous molecular dynamics-based methods for binding prediction for cellulases in a fraction of the time. Also, we show that when changed to a classification problem, the same network can be back-propagated to suggest mutations to improve enzyme binding. A similar approach could increase our understanding of the structure-activity relationship of enzymes, and suggest new promising mutations for enzyme design using explainable artificial intelligence.
\end{abstract}

\section{Introduction}
Cellulose is the most abundant organic compound on Earth and its utilization promises a way towards CO$\mathrm{_2}$-neutral fuels, materials, and chemicals.\cite{Ragauskas2006,Payne2015} However, it is very recalcitrant and hard to utilize. Over millennia, Nature developed a toolbox of enzymes to make use of this polymer. Primarily certain fungi and bacteria are known for their biomass-degrading capabilities. For simplicity, we will focus on the well studied and industrial very relevant fungal cellulases.\cite{Payne2015}
\begin{figure}
    \centering
    \includegraphics[trim=0 4cm 0 0,clip, width=\textwidth]{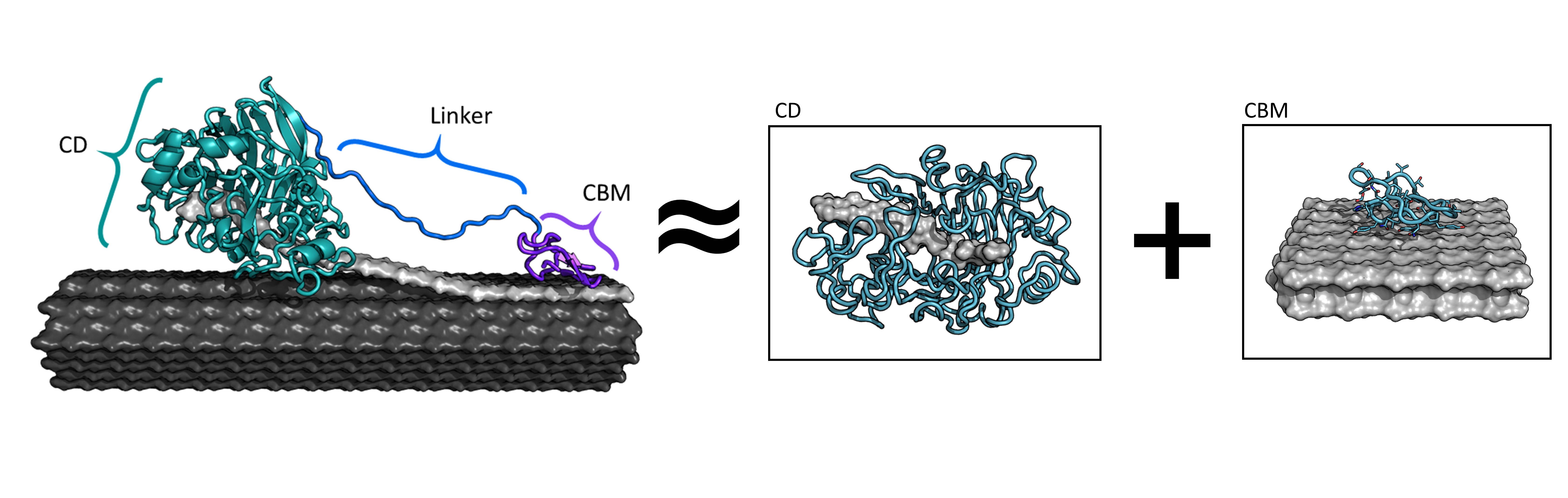}
    \caption{Illustration of the cellulase Cel7A from the fungus \textit{Trichoderma reesei} (\textit{Tr}Cel7A) bound on a cellulose fibril highlighting the typical domains of modular cellulases (left). The catalytic domain (CD) is bound to the carbohydrate-binding module (CBM) by a flexible linker.\cite{Payne2015} Previous work showed that the binding of the whole complex could be approximated by the binding of the CD towards a threaded chain and binding of the CBM on the crystalline surface (right).\cite{LFER1,LIE,VSGH7_Schaller2022s}}
    \label{fig:cbh_parts}
\end{figure}

Cellulases are typically modular, consisting of a catalytic domain (CD) and a carbohydrate-binding module (CBM) connected by a flexible linker (see Fig. \ref{fig:cbh_parts}).
The interactions between the enzymes and their substrate are very complex and non-trivial to set up for simulation. This complexity hinders computational exploration of these important enzymes compared to enzymes with a simpler mode of interaction. One important parameter to guide enzyme engineering is their binding affinity. However, computing of binding affinity \textit{in silico} is cumbersome and computationally expensive.

The enzyme-substrate complex of cellulases is mainly formed through interactions of the CD towards the threaded substrate chain and the CBM on the crystal surface. Earlier work showed that it can be approximated by the individual domains' binding to the substrate (see Fig. \ref{fig:cbh_parts}).\cite{LFER1,LIE,VSGH7_Schaller2022s} This simplifies the setup drastically, as crystal structures of the CD in complex with threaded substrate exists and the interactions of the CBM towards the crystal surface are well studied.\cite{Payne2015}

Building up on this approximation, we recently proposed a cheap computational method based on molecular dynamics (MD) simulation.\cite{LIE} While this method already has proven useful to screen through smaller enzyme families to discover new promising enzymes, it is still not feasible to compute properties of larger families or explore larger sequence space.\cite{VSGH7_Schaller2022s,VSGH6} Here, machine learning (ML) techniques promise to capture the complexity of the underlying interactions, providing reasonable binding estimates from either sequence or structure in an even shorter time.\cite{Greener2021,Mazurenko2019ML4Enz,Sally2020} Additionally, when at least partially based on real experimental results, machine learning could not only lead to faster predictions but also to more accurate ones than compared to the semi-empirical MD-based 2D-LIE screening method.\cite{Noe2020,LIE}

Cellulases can be found in different glycoside hydrolase (GH) families and have therefore different folds of the catalytic domain. While bacterial cellulases can possess CBMs from different CBM families, the here investigated fungal enzymes only have CBMs from a single family (CBM1).\cite{CAZY1,Payne2015} These different folds of the CDs imply that sequence and their related structural parts are not aligned across the families. However, as they act on the same substrate and consist of the same building blocks, their structure-function relationship should be common. This is also indicated by the capability of the semi-empirical 2D-LIE to predict across different families by utilizing common scaling parameters.\cite{LIE} Because of this, we focus on structure-based machine learning, which hopefully allows the model to easier transfer knowledge across families. Additionally, there are plenty of crystal structures available for the well-studied cellulases, which makes modeling of their structure very straightforward, even before the advent of \textit{Alphafold} (2.0).\cite{LIE,Jumper2021HighlyAlphaFold}


Within machine learning, there is a wide array of methods, but in this work, because we want to start from structure, methods originating from image recognition are well suited for the task. Especially, convolutional neural networks (CNN) have been shown to be capable of capturing underlying patterns in structural data, such as medical imaging data.\cite{CNN} Earlier work also showed that it is possible to generalize across different protein structures and predict protein-ligand binding via CNN.\cite{Jimenez2018,Dziubinska2018DevelopmentPrediction,Jones2021} However, those efforts focused on globular enzymes binding to small molecules. As cellulases are multi-domain enzymes binding to an interfacial substrate, those approaches are not readily transferable.

\section{Experimental procedures}
\subsection{Data set \& data curation}
\begin{landscape}
\begin{figure}
    \centering
\begin{minipage}{0.75\textwidth}
\centering
    \includegraphics[width=0.75\textwidth]{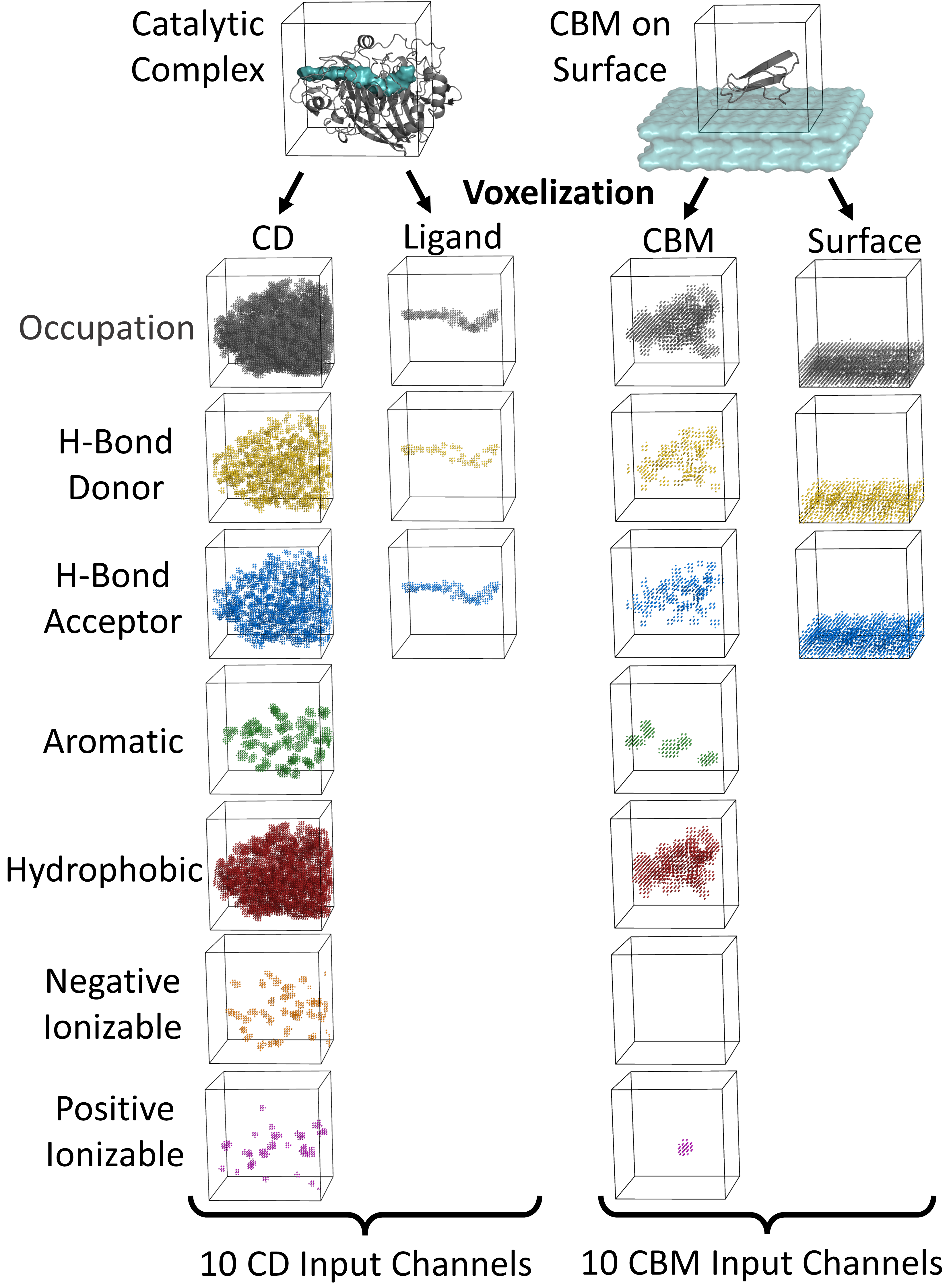}
    \captionof{figure}{Input example for the enzyme \textit{Tr}Cel7A. 10 channels were employed for both domains (CD with polymeric ligand and CBM on surface). 7 channels encode the protein (CD/CBM) and 3 encoding the organic molecules (polymeric ligand/surface). The CBM of \textit{Tr}Cel7A does not have negative ionizable atoms and the 6th input channel of that domain is therefore empty.}
    \label{fig:voxels}
\end{minipage}
\hspace{1cm}
\begin{minipage}{0.55\textwidth}
    \centering
    \includegraphics[width=0.85\textwidth]{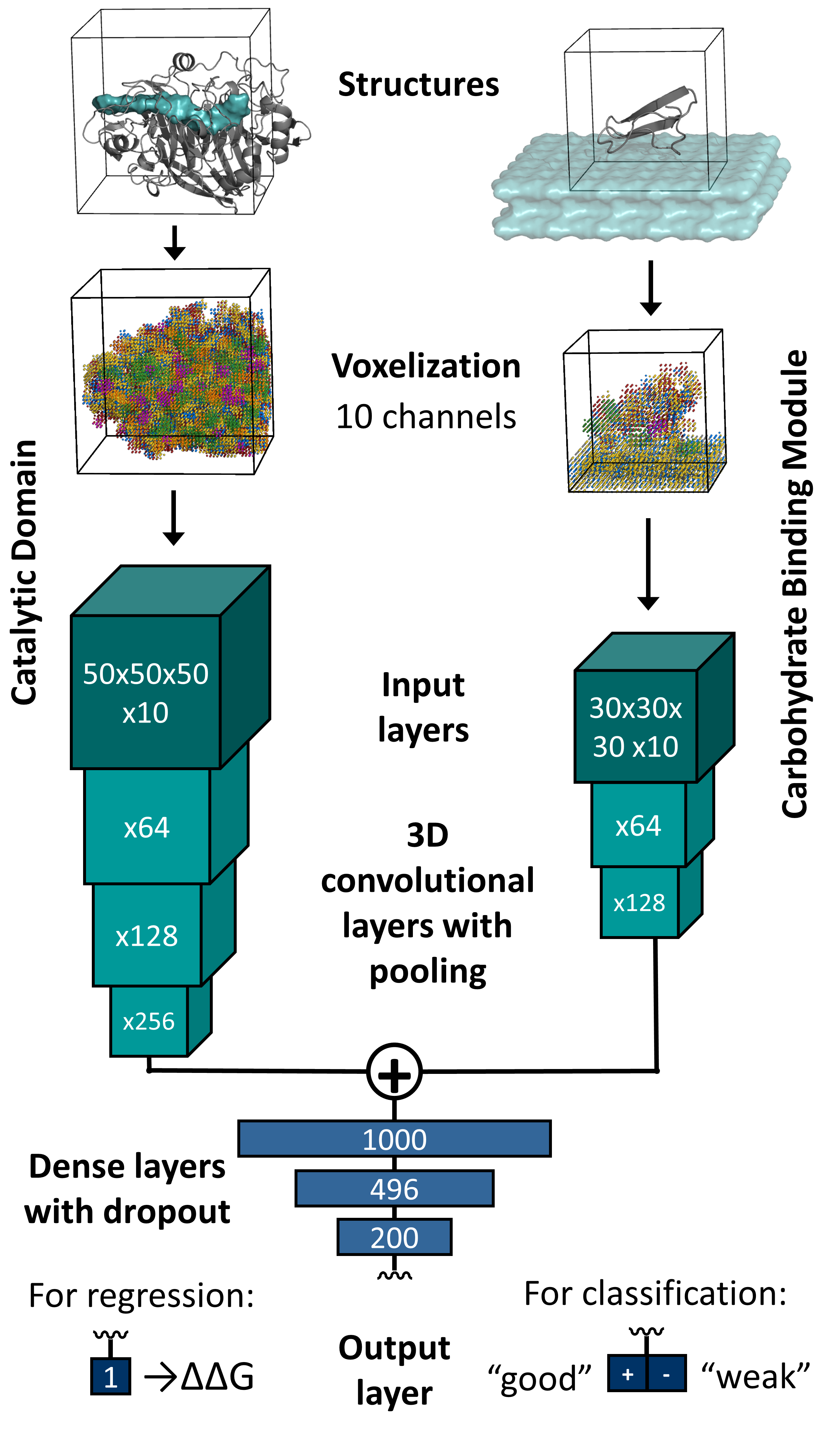}
    \captionof{figure}{Architecture of the dual-input CNN for binding prediction of multi-domain cellulases. For regression a single output node is used, while for classification 2 nodes were used to distinguish between "good" and "weak" binders.}
    \label{fig:cnn}
\end{minipage}
\end{figure}
\end{landscape}

In two recent studies, we performed virtual screenings with MD-based binding estimates for two cellulase families, GH6 and GH7.\cite{VSGH7_Schaller2022s, VSGH6} Family GH7 consisted of approx. 400 representative enzymes, family GH6 of about 1'200. Additionally, we earlier published a (relatively) large dataset of ca. 100 cellulases with experimental binding values.\cite{LFER1} For a more efficient training, all binding energies were linearly scaled with a minimum value of $-13$~kJ/mol and a maximum value of $17$~kJ/mol, resulting in values well between $0$-$1$ for all datasets.
The structures were aligned per domain type and split into protein and organic molecules. \texttt{Moleculekit} was used to voxelize the structures according to earlier work by Jimenez and co-workers with a resolution of $\mathrm{1}$~\AA.\cite{Jimenez2018} The original publication used 8 different channels. For the protein parts (CD/CBM), we dropped the "metal ion"-encoding channels as all of our homology-modeled structures are free of ions, thus resulting in 7 channels for the protein part of each domain (see Fig. \ref{fig:voxels}). For the organic molecule part (polymeric ligand/crystal surface), we additionally dropped the channels encoding aromatic, hydrophobic, positive ionizable, and negative ionizable atoms, because cellulose does not have atoms with those attributes. This led to 3 channels for the cellulose part of each domain. In total, this gave us 10 channels per domain, 7 from the protein part (CD/CBM) and 3 from the cellulase part (bound ligand/crystal surface, see Fig. \ref{fig:voxels}).

As inputs, a $\mathrm{50}$~\AA~cube from the voxelized CD data and a $\mathrm{30}$~\AA~from the CBM data was taken. In all cases, an $80\%/20\%$ train-test split was used. Enzyme families were proportionally stratified across the test and training set. To increase generalization and allow positional invariance, the training data was augmented. Random translations of the window taken as input cube in all three dimensions (up to 12~\AA~in all directions) were done. Additionally, random rotations for all possible 24 cube orientations, and random flips in all three dimensions were performed for the CD and CBM input. Accounting for permutations between the CD and CBM input, this results in millions of possible combinations. On-the-fly augmentation was implemented and tested but led to poor computational performance. Therefore, all datasets were split beforehand, and a fixed number of augmentation were done on the training data.
For training on the larger dataset based on GH6 and GH7, we generated $5$ augmented inputs per data point; for the smaller dataset based on the experimental dataset, we generated $20$ augmented inputs per data point.
Ideally speaking, even the larger simulated set is still regarded as small for deep learning applications, but biological data is costly to obtain. 

\subsection{Network}
The applied CNN architecture was inspired by literature.\cite{Jimenez2018,Dziubinska2018DevelopmentPrediction} Those literature models, however, only treat globular (single domain) enzymes. To be able to predict binding for our two-domain enzymes such as cellulases, we employ two input and two convolutional parts, one for each domain (see Fig. \ref{fig:cnn}).
The CD input, starting from a $\mathrm{50}$~\AA, was fed into three 3D-convolutional layers (64, 128, and 256 filters, respectively). The smaller CBM input was convoluted with two layers (64 and 128 filters, respectively). For all convolutional layers, a kernel size of 3 was used.  All convolutional layers were followed by maximum pooling layers. The convolutional parts were followed by a global maximum 3D-pooling layer, and a concatenation between CD and CBM feeds. 
This was followed by dense layers with 1000, 496, and 200 neurons (multiples of 8 for more efficient GPU training) with $\mathrm{ReLu}$ activation. For regression training, this was fed into a single output neuron with a $\mathrm{sigmoid}$ activation function. For classification, this was swapped with an output layer of $2$ neurons with $\mathrm{SoftMax}$ activation.

\subsection{Training}
Because we have datasets with different accuracy (simulated and experimental target data), we employed fine-tuning. Initially, we learned on the large dataset based on simulated data, and after that refined the model while learning on the more accurate but smaller set with experimental target values. Later, we used this pre-trained regression model and transferred it to a classification model for class-specific guided back-propagation. 

\paragraph{Initial Learning on Simulated Values}
Different hyperparameters were investigated. The final learning used stochastic gradient descent (SGD) with no momentum and a fixed learning rate of $10^{-4}$ for optimization.\cite{Bottou2012} Mean-squared error (MSE) was used for loss, and mean-absolute error (MAE) was tracked for performance checks. A batch size of $256$ with pre-fetching and interleaved file reading was used for all trainings. Batch normalization was applied to the convolutional layers, and a dropout of $0.6$ was used in the dense layers.

\paragraph{Fine-Tuning on Experimental Values}
Only a few things were changed compared to the initial learning: The learning rate was decreased to $10^{-5}$, and the dense layers were kept frozen during training. 

\paragraph{Transfer-Learning for Classification Model} 
The output layer was changed from a single regression neuron to $2$ neurons for classification.
An arbitrary energy limit was introduced at $0$~kJ/mol to split the dataset into a "good binder" and a "weak binder" class. 
Continuous target values were on-the-fly converted to sparse encoded vectors. Categorical cross-entropy loss and accuracy were used to guide the learning. The loss was class-weighted to account for data imbalance in the different classes. A slower learning rate of $10^{-6}$ was used while all dense layers were kept frozen.

\paragraph{Engineering Guided by Grad-CAM}
The trained classification model was taken, and gradient-weighted class activation mapping (Grad-CAM) was performed to investigate the activation of the input layers per class.\cite{GradCAM} \textit{Tr}Cel7A-$\mathrm{W38A}$ and \textit{Tr}Cel7A-$\mathrm{W40A}$, two known mutants of \textit{Tr}Cel7A were taken as a show-case. It is known that the residues $\mathrm{W38}$ and $\mathrm{W40}$ are critical for binding and that their mutation to alanine leads to a drastic decrease in binding energy.\cite{LFER1,Rojel2020}

\paragraph{Software}
The deep learning model was built with \texttt{Tensorflow} (2.4) using \texttt{python} (3.8.11), \texttt{CUDA} (11.3), \texttt{scikit-learn} (0.24.2).\cite{TENSORFLOW, SCIKIT} The module \texttt{3D-GuidedGradCAM} was used to perform guided back-propagation.\cite{GradCAM}

\section{Results}

\begin{figure}[p]
    \centering
    \vspace{-0.5cm}
    \includegraphics[width=\textwidth]{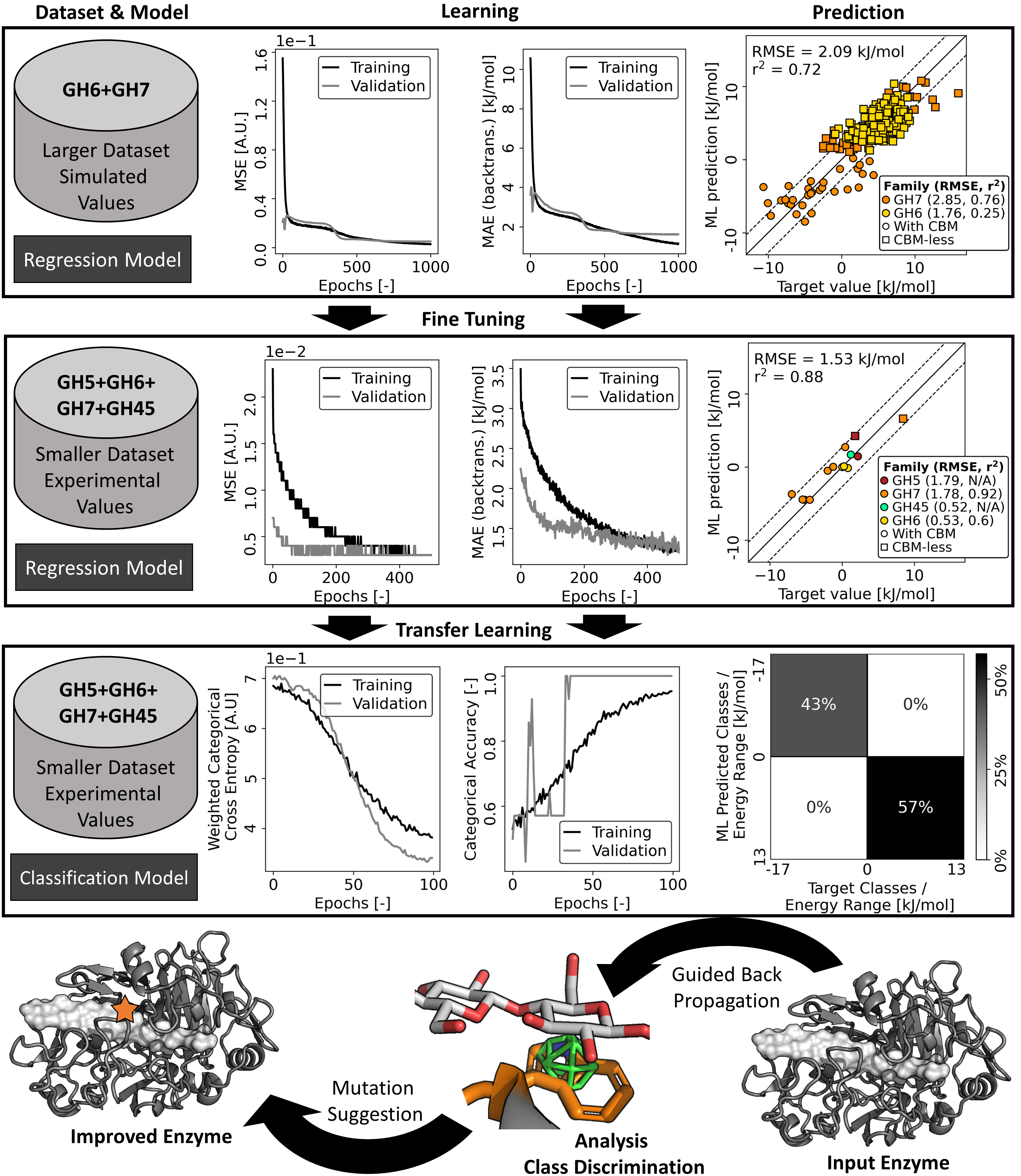}
    \caption{Overview over different learning stages with the used datasets, exemplary learning curves, and prediction plots for the validation set. In the prediction plots (right), the performance of the previous 2D-LIE method ($\pm$RMSE) is shown as dashed lines.\cite{LIE} Top: Initial learning of the regression model on the larger dataset with simulated values. Center: Fine-tuning of regression model on the smaller dataset with experimental values. Bottom: Transfer learning of the classification model based on the pre-trained regression model. Guided Grad-CAM can be performed, and subsequent class discriminant analysis can elucidate mutations for improved binding.}
    \label{fig:results}
\end{figure}
\paragraph{Initial Learning on Simulated Values} 
Learning on the larger dataset based on simulated target values was performed over 1000 epochs (see top of Fig. \ref{fig:results}). No increase in validation performance was observed at the end, suggesting that no overfitting has yet been reached. 
The final model resulted in a root-mean-squared error (RMSE) of $2.09$~kJ/mol, only slightly higher than the estimated $1.91$~kJ/mol for the initial simulation method at longer simulation times ($10$~ns instead of the $1$~ns, that was used for the generation of the dataset), which could be close to the accuracy limit of the dataset.
\paragraph{Fine Tuning on Experimental Values}
Learning was performed over 500 epochs (see center of Fig. \ref{fig:results}). Again, no increase in validation performance was observed at the end, suggesting that overfitting has not been reached yet. The final model resulted in an RMSE of $1.53$~kJ/mol, outperforming the previous method 2D-LIE (RMSE of $1.98$~kJ/mol). 
This performance is comparable to related methods for globular enzymes from literature, despite predicting binding of more complex modular enzymes. \cite{Dziubinska2018DevelopmentPrediction,Jimenez2018,Jones2021,Sally2020}
The regression model seems to work well for all included families (GH5, GH6, GH7, and GH45) and irrespective of modularity ($\pm$CBM).
\paragraph{Transfer-Learning for Classification Model}
Learning was performed over 100 epochs leading to a perfect categorical accuracy on the validation set (see bottom of Fig. \ref{fig:results}). The jumps in the accuracy suggest that using a different, more stable optimizer for the training could be advantageous.

\begin{figure}
    \centering
    \includegraphics[width=\textwidth]{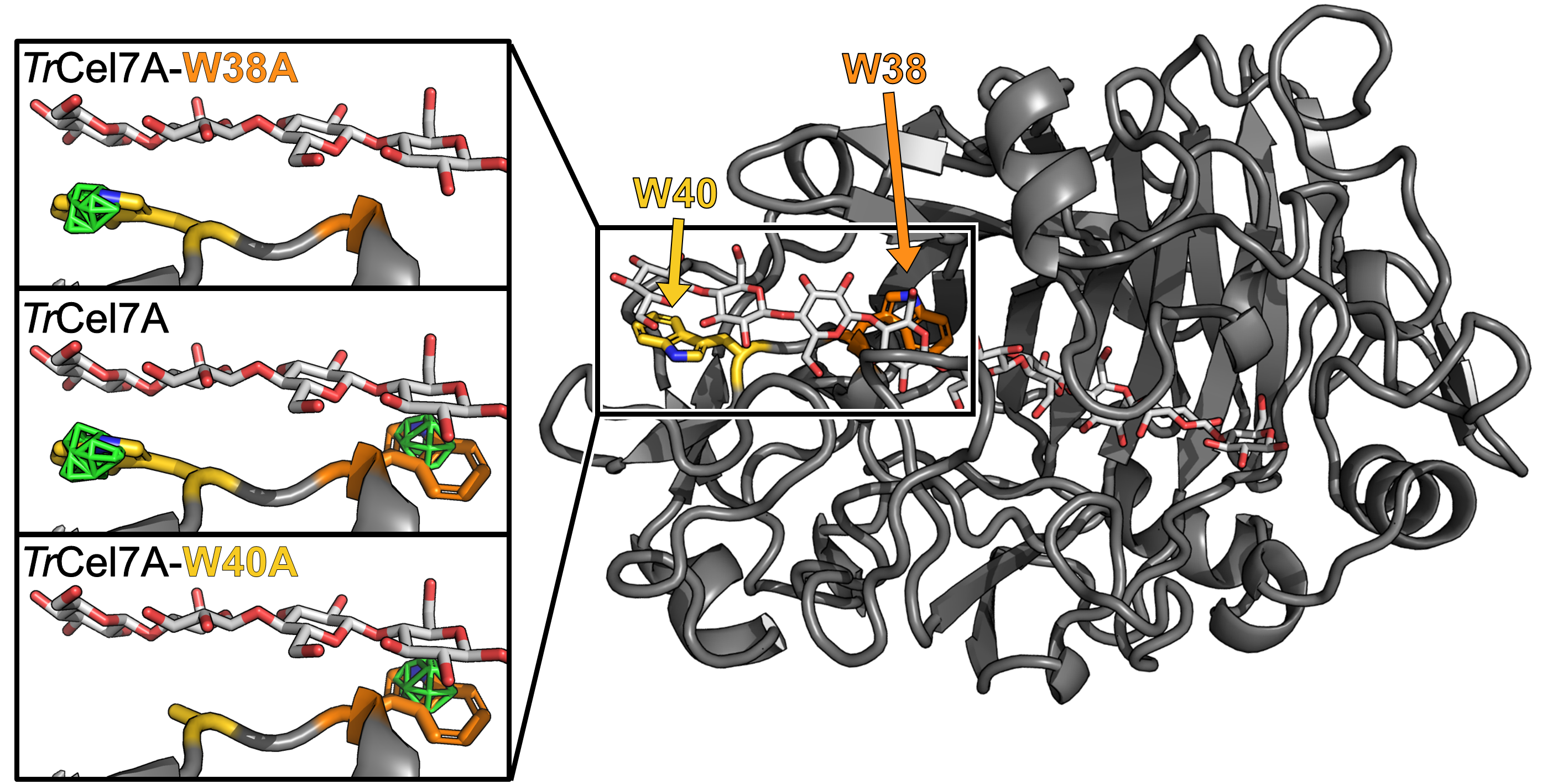}
    \caption{Class discriminant analysis of two \textit{Tr}Cel7A variants (W38A and W40A) with Grad-CAM. Position 38 is highlighted in orange, position 40 in gold. Both residues are found in the tunnel entrance (right) and are known to stack with the threaded cellulose chain. Inlets (left) show "good binder" class activation (top 25\%) as green isomeshes for the wildtype \textit{Tr}Cel7A and the two variants. The presence of the tryptophan residue leads to class-specific activation, as seen by the appearance of the isomeshes.}
    \label{fig:backprop}
\end{figure}
\paragraph{Engineering Guided by Grad-CAM}
The wildtype (\textit{Tr}Cel7A) is found in the "good binder" class ($<0$~kJ/mol), both according to the ground truth and predicted by the classification model. The point mutations are not impactful enough to change the class, however, the change in activation can still be interpreted. \textit{Tr}Cel7A, \textit{Tr}Cel7A-$\mathrm{W38A}$, and \textit{Tr}Cel7A-$\mathrm{W40A}$ were inferred onto the model, and Grad-CAM was used to perform class discriminant analysis of the activation (see Fig. \ref{fig:backprop}). The presence of the tryptophans leads to an increased activation for the "good binder" class. In other words, the model becomes more sure that the enzyme binds well.


\section{Discussion}
The structure-activity relationship of proteins and their substrate is non-linear and very complex in general. Thus, they pose an ideal target for machine learning if a feasible encoding for the problem can be found. One possible approach is the voxelization of the structure. Ideally, this readily allows to capture fold-independent properties as it is solely structure and not sequence-based.\cite{Jimenez2018,Greener2021} In this work, we presented machine-learning models for multi-domain enzymes in general and cellulases in particular based on voxelized structures as inputs.
Cellulases are an important class of enzymes relevant for the conversion of biomass to fuels and chemicals. Their redesign is relevant for improved activity under industrial conditions, and one important descriptor for their engineering is their binding affinity. However, the computational assessment of their binding is non-trivial as their enzyme-substrate interaction is very complex. Therefore, we developed a method to predict their binding affinity to guide future enzyme engineering.

We used simulated binding data from two previous studies of family GH6 and GH7 to pre-train a regression model.\cite{VSGH6,VSGH7_Schaller2022s} Subsequently, we used a smaller set of experimental values\cite{LIE} to fine-tune the regression model, resulting in a model with an estimated RMSE of $1.53$~kJ/mol, outperforming previous computational screening methods for cellulases. Once trained, this regression model is also orders of magnitude faster than the previous MD-based method.
The trained regression models were transferred to a classification model to allow class-specific guided back-propagation. The classification model itself is not that practical, but it allows to use of explainable artificial intelligence (XAI) methods developed for classification problems.\cite{Streib2020_XAIpersp,Angelov2021_XAIrev,Confalonieri2021_XAIhist} Two mutant enzymes with poor binding were class-specific back-propagated, and the found densities suggested reintroducing the mutationally removed aromatic residues. While this reversal to increase binding once again is trivial, the same back-propagation approach could be used to find other unknown mutational suggestions to improve binding. Future work could directly use networks with simultaneous regression and classification output nodes to avoid retraining and additional hyper-parameter search. More general, regression and classification models like these could be investigated with additional XAI methods to learn more about the structure-activity relationship of enzymes. Approaches like these could make it feasible to search vast sequence space for interfacial enzymes such as cellulases and therefore push enzyme discovery and design to new limits. Using XAI methods could allow us to understand more of the hidden structure-activity relationship within the ML models and lead to a new understanding of enzyme-substrate interaction.

\subsection{Data availability}
All input data is based on public datasets as described in the referenced papers. All employed software is open-source and all necessary information is listed.

\subsection{Conflict of interest}
K.B. works for Novozymes A/S, a major enzyme-producing company.


\subsection{Acknowledgments}
The simulations were carried out at the high-performance cluster at the Technical University of Denmark.
This work was supported by the Independent Research Fund Denmark (Grant No. 8022-00165B) and the Novo Nordisk Foundation (Grant Nos. NNF15OC0016606 and NNF17SA0028392).

%
%

\bibliography{references}

\end{document}